\documentclass[aps,prl,twocolumn,superscriptaddress,amsmath,showpacs]{revtex4-1}

\usepackage{graphicx}
\usepackage{amsmath}
\usepackage{amssymb}

\begin{document}

\title{Deterministic Super Resolution with Coherent States at the Shot Noise Limit} 

% Place the author information here.  Please hand-code the contact
% information and notecalls; do *not* use \footnote commands.  Let the
% author contact information appear immediately below the author names
% as shown.  We would also prefer that you don't change the type-size
% settings shown here.

\author{Emanuele Distante$^{1,2}$}
\noaffiliation

\author{Miroslav Je{\v{z}}ek$^{1,3}$} 
\noaffiliation

\author{Ulrik L. Andersen$^{1}$}
\noaffiliation

\affiliation{Department of Physics, Technical University of Denmark, Fysikvej, 2800 Kgs. Lyngby, Denmark}
\affiliation{Dipartimento di Fisica, Universita degli Studi di Milano, I-20133 Milano, Italy}
\affiliation{Department of Optics, Faculty of Science, Palack{\'{y}} University, 17. listopadu 12, 77146 Olomouc}

% Include the date command, but leave its argument blank.

\date{\today}

%%%%%%%%%%%%%%%%% END OF PREAMBLE %%%%%%%%%%%%%%%%

\begin{abstract}
Interference of light fields plays an important role in various high-precision measurement schemes. It has been shown that super resolving phase measurements beyond the standard coherent state limit can be obtained either by using maximally entangled multi-particle states of light or using complex detection approaches. 
%In addition to their high technical complexity, these methods lack robustness or efficiency rendering the sensitivity performance above the shot noise limit. 
Here we show that super resolving phase measurements at the shot noise limit can be achieved without resorting to non-classical optical states or to low-efficiency detection processes. Using robust coherent states of light, high-efficiency homodyne detection and a deterministic binarization processing technique, we show a narrowing of the interference fringes that scales with $1/\sqrt{N}$ where $N$ is the mean number of photons of the coherent state. Experimentally we demonstrate a 12-fold narrowing at the shot noise limit. 
%Due to its immense simplicity, robustness and high performance, this method is likely to play an important role in future real-world implementations of high-resolution interferometers.   
\end{abstract}

%\baselineskip24pt

%\baselineskip24pt
\maketitle

When two coherent electromagnetic waves interfere as in Young's double slit
experiment or in a standard Mach-Zehnder interferometer, an oscillatory interference pattern arises with a periodicity
governed by the wavelength, $\lambda$, of the field. The period is given by $\lambda/2$ and is often referred to as the standard resolution limit of interferometers (and in imaging it is the Rayleigh resolution criterion~\cite{BornWolf}). 
%where $\lambda$ is the wavelength.
%This is a manifestation of Rayleigh's famous diffraction criterion stating
%that the minimal resolvable structure in lithography, microscopy and
%imaging is $\Delta x=\lambda/2$ \cite{BornWolf}.
% \cite{Scully2003,Zubairy2004}. 
Super resolution ---that is resolution beyond the standard $\lambda/2$ limit ---can be attained by the use of quantum entanglement.
E.g. using the maximally path-entangled multi-particle NOON states,
$|NOON\rangle=(|N0\rangle+|0N\rangle)/\sqrt{2}$, it is possible to
achieve super-resolution with resolvable features down to $\lambda/(2N)$
where $N$ is the number of photons \cite{Boto2000}. Super resolution
with NOON states has been demonstrated with ions \cite{Wineland2005},
nuclear spins \cite{Morton2009}, atoms \cite{Chen2010} and up to
four photons \cite{Rarity1990,Steinberg2004,Zeilinger2004,Takeuchi2007}.
In addition to super resolution, the NOON states can in principle
also beat the quantum shot noise limit (SNL) in phase estimation
ultimately reaching the optimal estimation known as the Heisenberg limit~\cite{giovanetti2011}. However, since these states
are extremely fragile and are prepared and detected with very low
efficiency \cite{Fiurasek2002}, it is experimentally very challenging
to beat the SNL \cite{Rubin2007,Walmsley2009}.

Coherent states of light have also been used to obtain super resolution. 
The idea is to detect a nonclassical state (such as the NOON state)
via state projection as opposed to nonclassical state preparation~\cite{Pregnell2004}. 
Examples of projections of coherent states that lead to phase
super resolution are photon counting, coincidence counting and
parity detection 
%and feedback controlled homodyne detection
~\cite{Boyd2004,Bouwmeester2006,White2007,
Silberberg2010,Dowling2010,Zappa2010,Kothe2010,Plick2010,Boyd}. 
Despite its super resolving capability, this state projection method cannot beat the SNL
in phase estimation but it may approach it for
an optimized parity detector~\cite{Dowling2010,Plick2010}. Although this method largely reduces the
complexity of the preparation stage, the detection part remains complex
(ideally requiring photon number resolving detectors) and the efficiency
in projecting out the desired non-classical state is often very low~\cite{White2007}.

%A common trait of the two methods is that their Wigner functions
%(quasi probability distributions in phase space) associated with
%the non-classical state preparation and state detection are
%non-Gaussian and negative. Recently, it was realized that the parity detector could be substituted with a homodyne detector combined with a fast feedback control system~\cite{Plick2010}.  
%(Such features have been shown to be
%a necessary condition for many quantum information protocols
%such as entanglement distillation and quantum computing).
In this Letter we propose and demonstrate a simple and very efficient
scheme to obtain super-resolution at the SNL without the need of
complex states in preparation or complex projectors in detection. 
We use coherent states of light and a simple high-efficiency homodyne detector to achieve
phase super resolution beyond what has been achieved with any
non-Gaussian resources or detectors. The method is deterministic and we show that it operates close to the shot noise limit in contrast to all previous implementations of coherent state based super-resolution.   

%In this Letter we show that super resolving phase measurements at the shot noise limit can be achieved without resorting to complex non-classical optical states or to low-efficiency and complex detection processes. Using robust coherent states of light, high-efficiency homodyne detection and a deterministic binarization processing technique, we beat the diffraction limit by 12 times at the quantum shot noise limit.

\begin{figure}[ht!]
	\centering	
	\includegraphics[width={1\linewidth}]{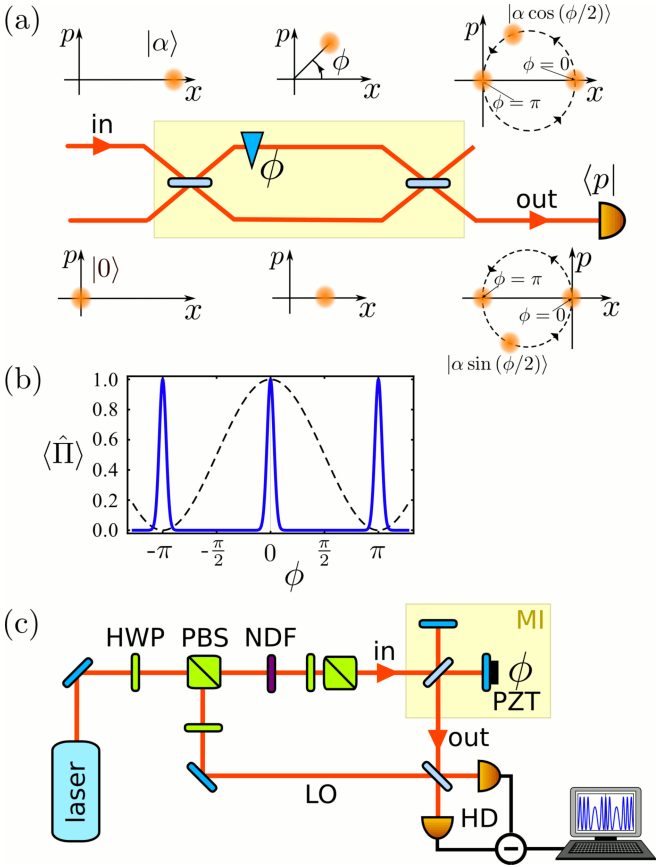}
	\caption{Experimental setup and illustration of the principles. 
	(a) Schematic of the experimental setup. A product of a coherent state, $|\alpha\rangle$, and a vacuum state, $|0\rangle$, is transformed through an interferometer and measured with a homodyne detector described by the ideal projector $\langle p|$. The evolutions in phase space of the two states are illustated by the insets.  	
	(b) The phase response function $\langle \Pi\rangle$ for the standard interferometer scheme (dashed curve) and for the super resolving scheme (solid curve).  
	(c) Experimental setup. Laser light is controllably divided into two beams using a half-wave plate (HWP) and a polarizing beam splitter (PBS) thereby creating a signal and a local oscillator (LO) beam for homodyne detection (HD). The power of the signal beam is controlled by a neutral density filter (NDF), a HWP and a PBS, and subsequently sent into a Michelson interferomter (MI), the function of which is identical to the one in (a). A piezo-crystal (PZT) attached to one of the interferometer mirrors scans the phase, $\phi$, and the resulting output is measured by means of HD.       
  }
  \label{figure1}
\end{figure}

Our method follows the interferometric scheme illustrated
in Fig.~\ref{figure1} (a). A coherent state of light, $|\alpha\rangle$,
with amplitude $\alpha$ and mean photon number $N=|\alpha|^2$ enters
the interferometer at the input symmetric beam splitter.
The resulting state, $|\frac{\alpha}{\sqrt{2}}\rangle|\frac{\alpha}{\sqrt{2}}\rangle$,
acquires a phase shift, $\phi$, in one arm of the interferometer,
and the final state at the output,
$|\Psi(\phi)\rangle=|\cos(\phi/2)\alpha\rangle|\sin(\phi/2)\alpha\rangle$,
is produced by interference at the second beam splitter.
The observed phase resolution and sensitivity crucially depend
on how this output state is detected. By describing the detection
process with a measurement operator $\hat{\Pi}$, the response
function of the setup is the mean value of that operator,
$\langle\hat{\Pi}\rangle$. Using for example a standard intensity
detector described by the observable $\hat{\Pi}=\hat N$, the detector response
is $N\cos^2(\phi/2)$ which is an oscillating function with a period
given by $\lambda/2$, thus coinciding with the standard resolution limit. In the following we show that by substituting the intensity detector with
a simple homodyne detector combined with a post binarization process, we beat the standard resolution limit.

%A homodyne detector (HD) measures the eigenvalues of a quadrature operator
%$\hat{x}\cos\theta + \hat{p}\sin\theta$, where $\theta$ is the phase of
%the local oscillator intrinsic to the HD. $\hat{x}$ and $\hat{p}$
%are the canonically conjugated amplitude and phase quadrature operators
%related to the bosonic field operator; $\hat{a} = \hat{x} + {\rm i}\hat{p}$, and
%obeys the commutation relation $[\hat{x},\hat{p}]={\rm i}/2$. 

The main idea of our approach is to divide the quadrature measurement
outcomes into two bins: Assuming that we measure the phase quadrature, $\hat{p}$, we classify
two different results associated with outcomes in the intervals
$|p|\leq a$ and $|p|>a$. Such a measurement strategy is described
by the two projectors
\begin{equation} \label{POVM}
  \hat{\Pi}_0 = \int_{-a}^{a} \!\! {\rm d}p \, |p\rangle\langle p|,
  \qquad
  \hat{\Pi}_1 = \hat{{\rm I}} - \hat{\Pi}_0
\end{equation}
and the measurement observable can thus be written as $\hat{\Pi} = \lambda_0 \hat{\Pi}_0+\lambda_1 \hat{\Pi}_1$ 
%\begin{equation} \label{detection}
%  \hat{\Pi} = \sum_{k=1,2}\lambda_k \hat{\Pi}_k,
%\end{equation}
where $\lambda_0 = 1/{\mathrm{erf}}(\sqrt{2}a)$ and $\lambda_1=0$
are the eigenvalues associated with the two measurement outcomes. 
%The detection procedure returns either $\lambda_0$ for $p\in[-a,a]$
%or $\lambda_1$ elsewhere and yields the response function
The detector response function of this dichotomic
strategy is 
\begin{equation} \label{response}
\begin{array}{c}
  \displaystyle{
  \langle\hat{\Pi}\rangle
  \displaystyle{
  = \frac{1}{\mathrm{erf}(\sqrt{2}a)}\int_{-a}^{a} \!\!\mathrm{d}p
  \, |\langle \sqrt{N} \cos(\phi/2)|p\rangle|^2. }}
\end{array}
\end{equation}
For a general value of $a$ the response function cannot be evaluated
in closed form in terms of elementary functions
% $f_a(\phi)=\frac{1}{2\mathrm erf (\sqrt{2}a)}\left\{\mathrm{erf}\left[
% \sqrt{2}\left(a-\frac{1}{2}\sqrt{N}\mathrm{sin}\,\phi\right)\right]+\mathrm{erf}
% \left[\sqrt{2}\left(a+\frac{1}{2}\sqrt{N}\mathrm{sin}\,\phi\right)\right]\right\}$
but for $a\rightarrow 0$ (corresponding to binning the results for which
$p=0$ and $p\neq0$), it can be simply written as
\begin{equation} \label{response0}
  \langle\hat{\Pi}\rangle = \exp{\left(-\frac12 N\sin^2\phi\right)}
\end{equation}
and is illustrated in Fig.~\ref{figure1} (b).
The full width at half maximum of this fringe is
${\rm FWHM} = 2\arcsin{ \sqrt{(2\ln2)/N} }$,
and by comparing it to the FWHM of the fringe associated with
a standard resolution limited intensity detection system,
we see that super-resolution is obtained
for $N>2\ln{2}$. For $N\rightarrow\infty$ we find a $1/\sqrt{N}$
improvement of the resolution with respect to the standard limit.    
%For the more relevant case of $a>0$, the 
%The same scaling is attained for $a>0$ as shown in~\cite{SM}.

In addition to being super-resolving, our approach also exhibits
a phase sensitivity at the SNL. The sensitivity is defined as
  $\Delta\phi = \Delta{\Pi} / \left|
  \rm{d}\langle\hat{\Pi}\rangle/\rm{d}\phi \right|$
% $\Delta\phi = \Delta\hat{\Pi} / \left|
% \frac{\rm{d}\langle\hat{\Pi}\rangle}{\rm{d}\phi} \right|$
where $\Delta{\Pi}=\sqrt{\langle\hat\Pi^2\rangle-\langle\hat\Pi\rangle^2}$, and for our measurement operator (with $a\rightarrow 0$) it reaches
the minimum value of 
%$\Delta\phi_{\rm min} \approx \frac{1.03}{\sqrt{N}}$
\begin{equation}\label{sensitivity0_min}
	\Delta\phi_{\rm min}=\sqrt{\,\sqrt{\frac{\pi}{2}}\,\frac{e^{\frac{1}{4}\left(2+N-\sqrt{4+N^2}\right)}-\sqrt{2/\pi}}{\sqrt{4+N^2}-2}}
\end{equation}
near the phase points $\phi_{\rm min}=\pm\,\mathrm{arccos}\left(\sqrt{1/2}-1/N+\sqrt{4+N^2}/2N\right)$
and for large $N$ it converges to 
$\Delta\phi_{\rm min} =
 \sqrt{\left(\sqrt{{\rm e} \pi /2}-1\right)/N}\approx 1.03/\sqrt{N}$,
thus being close to the SNL. 
%This feature, preserved
%even for a finite value of $a$ parameter, assures our scheme
%works at quantum shot noise level and the results are not obscured
%by classical noise.

\begin{figure}[ht!]
	\centering	
	\includegraphics[width=1\linewidth]{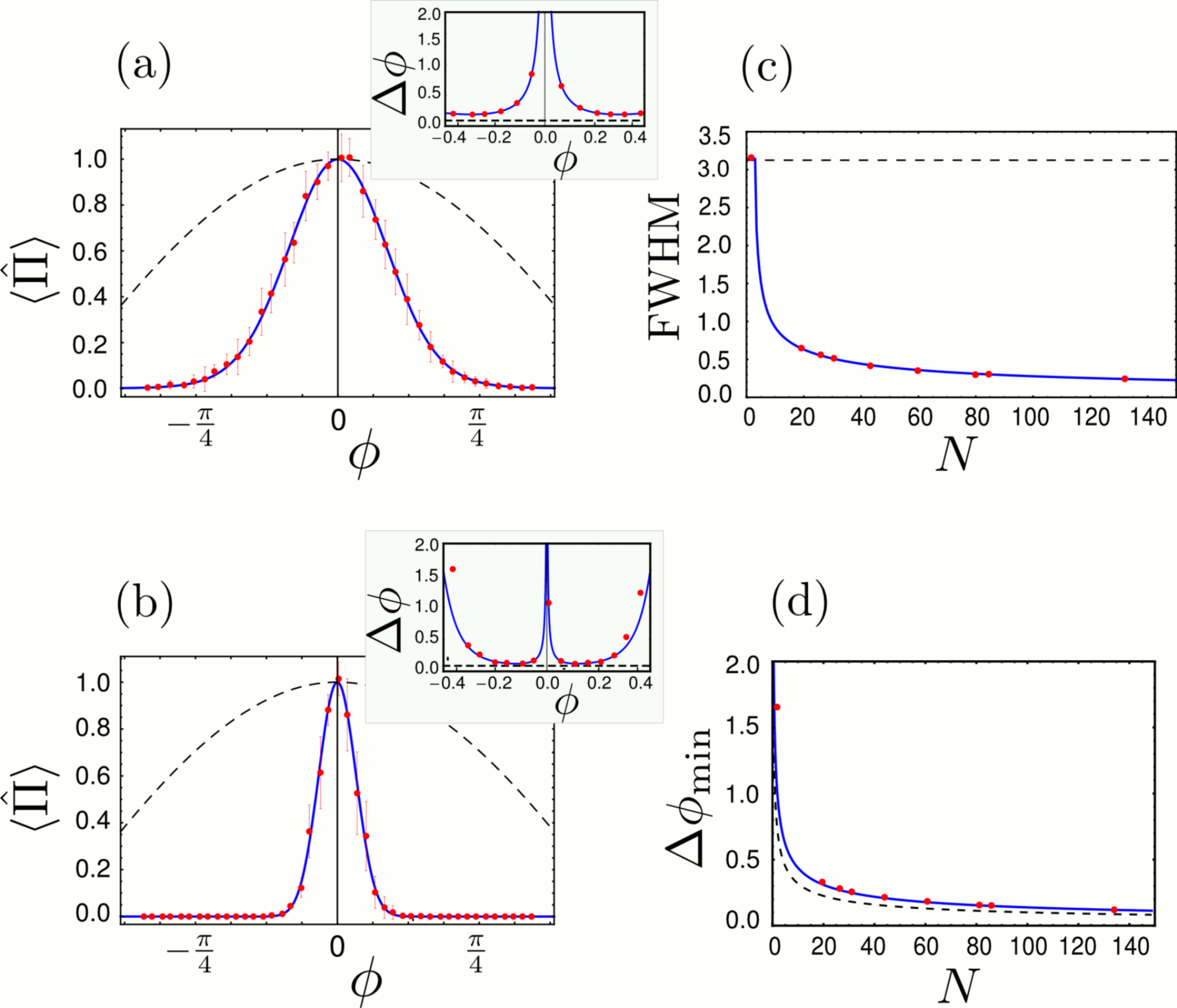}
	\caption{ Performance of the super resolving interferomenter.
	(a) and (b) show the experimental results (dots) and the theory (solid curve) for the response
	function with $a=0.5$ for two different mean photon numbers
	$N=19$ and $N=132$, respectively. The dashed curves represent the standard Rayleigh limited
	strategy. The insets of (a) and (b) are
	the results of the sensitivity as a function of $\phi$ for the same values of $N$.
	The dashed curves represent the SNL.
	(c) The resolution in terms of FWHM of our scheme
	as a function of the mean photon number (red dots) together with theory for our approach (solid curve) and for Rayleigh limited startegy (dashed line). 
	%Blue line represents the
  %theoretical curve $FWHM\propto1/\sqrt{N}$
  %while the dashed line shows the minimum theoretical FWHM
	%of a standard intensity detection system.
	%
	(d) Minimum value of the sensitivity for our scheme for different mean photon numbers (red dots).The solid curve represents the theoretical curve, $\Delta\phi_{min}\approx1.37/\sqrt{N}$, and the dashed line stands for the SNL.
	}
  \label{figure2}
\end{figure}

In the limit of $a\rightarrow0$ the measurement is not physically sound
as it requires infinitely high energy. However, both the resolution and
sensitivity properties are preserved even for a finite value of $a$.
Indeed, for a general value of $a$ the response function can be evaluate as
\begin{eqnarray} \label{response}
  \langle\hat{\Pi}\rangle=\frac{1}{2\mathrm{erf}(\sqrt{2}a)}\left\{\mathrm{erf}\left[\sqrt{2}g_+\right]+\mathrm{erf}
  \left[\sqrt{2}g_-\right]\right\},
\end{eqnarray}
where $g_\pm=\sqrt{2}\left(a\pm\frac{1}{2}\sqrt{N}\mathrm{sin}\,\phi\right)$. The response function and its width is illustrated by the solid curves in Fig.~2 (a)-(c). The scaling of the width is again found to be $1/\sqrt{N}$. Finally, the sensitivity for a finite $a$ reads
\begin{equation}\label{sensitivity} 
		\delta\phi = \sqrt{\frac{\pi}{2}\frac{e^{\left(2a+\sqrt{N}\mathrm{sin}\,\phi\right)^2(2-k)k}}{N\,\mathrm{cos}^2\,\phi\,\left(e^{4a\sqrt{N}\mathrm{Sin}\,\phi}-1\right)^2}}
\end{equation}
where $k=\mathrm{erfc}\left[\sqrt{2}g_-\right]+\mathrm{erfc}\left[\sqrt{2}g_+\right]$ and $\mathrm{erfc}(\cdot)=1-\mathrm{erf}(\cdot)$ is the complementary error function.
For $a=1/2$ and $N>>1$ the sensitivity follows the shot noise scaling; $\Delta\phi_{\rm min} \approx \frac{1.37}{\sqrt{N}}$.
The value of $a=1/2$ represents a trade-off between sensitivity and resolution.
For example, for higher value of $a$ the sensitivity is closer to the SNL
but the resolution FWHM increases and vice versa.

We now implement the protocol using the setup displayed in Fig.~\ref{figure1} (c). 
A coherent state with a controllable mean photon number is sent through a Michelson interferometer (MI) in which the relative phase ($\phi$) is continuously varied by a piezo driven mirror. The phase quadrature of the interferometer output state is then measured with a high-efficiency homodyne detector (HD). We subject the resulting detector outcomes to the binning procedure with the interval set to $a=1/2$. 
The experiment is repeated several times and the frequencies at which the measurement outcomes fall within the two quadrature intervals (described by the projectors (\ref{POVM})) are found for different relative phases, and the resulting response functions are plotted in Fig.\ref{figure2} (a)-(b) for two different power levels (red dots).
A clear narrowing of the fringe with respect to the fringe for the standard approach (dashed curve) is observed, thus proving the super resolution capabilities of our scheme. We also note that the visibility of the new interference fringes is basically unchanged and close to unity. We repeat the experiment for several different mean photon numbers and the results of the FWHM are summarized in Fig.\ref{figure2} (c).

%The visibility of the interference fringes remains high indicating that the phase sensitivity (partially given by the slope of the fringe) is kept at the SNL. 
We also estimate the sensitivity from the measurements and the results for two different mean photon numbers are presented by the insets of Fig.~\ref{figure2}. These results  demonstrate that the measurements possess a phase sensitivity very close to the SNL (dashed line) for certain phases. In Fig.~\ref{figure2} (d) we present a summary of the optimal phase sensitivities for several different mean photon numbers and compare it to the SNL (dashed curve).

As we have now seen, binary binning of quadrature measurements leads to a narrowing of the interference fringe. However, the number of fringes in an $2\pi$ period remains unchanged as opposed to interferometry with NOON states where the number of fringes increases with the photon number. It is, however, also possible with coherent states to increase the number of fringes in a period by employing a multiple binning approach: Instead of dividing the measurement results in two different intervals, we divide them into multiple intervals consisting of equidistant bins with length $2a$. This is described by the projectors
\begin{equation} \label{pi0 real multi:eq}
	\hat{\Pi}_k = \int_{b_k-a}^{\,b_k+a}|p'\rangle\langle p'|\mathrm{d}p'
	\qquad
	\hat{\Pi}_{n+1} = \hat{\rm I}-\sum_{k=1}^{n} \hat{\Pi}_k
\end{equation}
By setting the eigenvalues $\lambda_k=1/\mathrm{erf}(\sqrt{2}a)$
for $k\in\{1,\ldots,n\}$ and $\lambda_{n+1}=0$, we find the resulting response function
\begin{eqnarray} \label{out multi real:eq}
 \langle\hat{\Pi}\rangle = \displaystyle \frac{1}{2\mathrm{erf}(\sqrt{2}a)} \,\sum_{k=0}^{n+1}
	\left\{\mathrm{erf}\left[\sqrt{2}\left(g_-b_k\right)\right]+\mathrm{erf}\left[\sqrt{2}\left(g_++b_k\right)\right]\right\} & \nonumber
\end{eqnarray}
where $b_k$ is the central position of the intervals on the p-quadrature line. If the distance between the intervals is $b>2a$, the projectors are orthogonal and we can straightforwardly find the quantity
%\begin{eqnarray}
%\langle\hat{\Pi}\rangle=\frac{1}{{\rm erf}(\sqrt{2}a)}\sum_{k=1}^{k=n}\langle\hat{\Pi}_k\rangle\\
%\langle\Delta\hat{\Pi}^2\rangle=\langle \hat{\Pi}\rangle\left(\frac{1}{{\rm erf}(\sqrt{2}a)}-\langle\hat{\Pi}\rangle\right)
%\end{eqnarray}
\begin{equation}
\langle\Delta\hat{\Pi}^2\rangle=\langle \hat{\Pi}\rangle\left(\frac{1}{{\rm erf}(\sqrt{2}a)}-\langle\hat{\Pi}\rangle\right)
\end{equation}
and thus the sensitivity which is plotted in Fig.~3(b) (solid curve).

\begin{figure}[ht!]
	\centering	
	\includegraphics[width=1\linewidth]{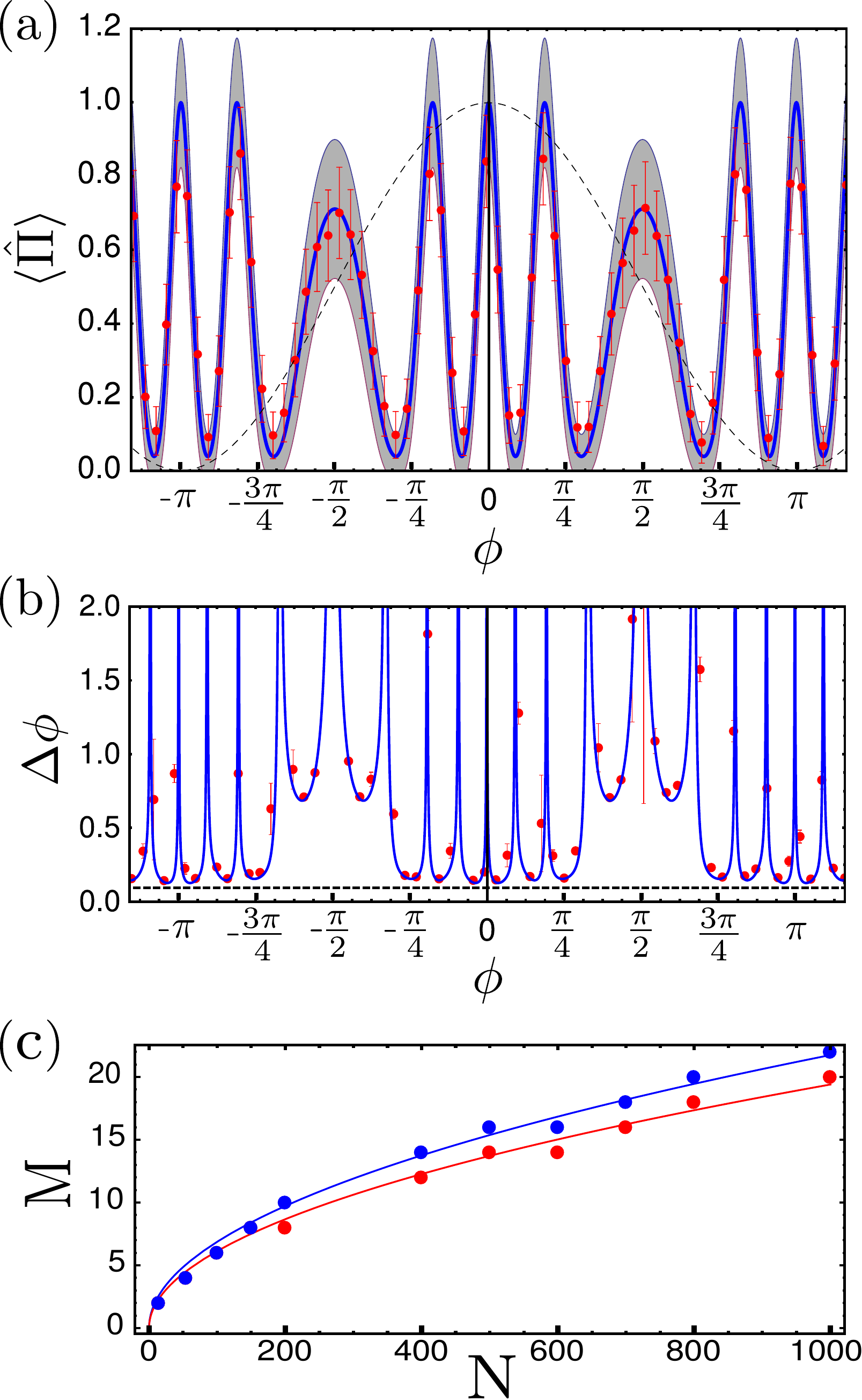}
	\caption{ Results for the multiple binning approach.
	(a) Response function with 8 fringes per period obtained by binning the measurement outcomes along 5 intervals for a coherent state with $N=139$. The central points of these intervals are located at $p=b\cdot k$ where $k\in[-2,-1,0,1,2]$ and $b=3.17$. The average fidelity of all fringes within a period is 95\%. The data (red dots) fit
	well with theory (blue line) and the uncertainty for each
	point lies inside the theoretically predicted uncertainty,
	represented by the shaded area. Interference fringe corresponding to the Rayleigh limited approach is shown by the dashed curve. 
	(b) Phase sensitivity for the multiple binning approach associated with the experimental results (dots) and the theory (solid curve). Near SNL performance (represented by the dashed line) is obtained for several phases. (c) Plot of the number of fringes $M$ for which the visibility is larger than $0.95$ (red) and $0.90$ (blue) as a function of the average photon number, $N$, of the coherent state. The dots correspond to numerical estimates whereas the solid curves are theoretical fits.}
  \label{figure3}
\end{figure}

An example is shown in Fig.~\ref{figure3} for $a=1/2$ and 8 fringes. The multiple fringe appproach will, however, give rise to a slightly lower visibility of the interference fringes, thus rendering a trade-off between the number of fringes within a period and the visibility of the resulting pattern for a given average photon number. It is however possible to recover the near unity visibility by increasing the number of photons as illustrated in Fig.~\ref{figure3}c.  
%Keeping the number of fringes linearly proportional to $\sqrt{N}$ the visibility remains constant independent of $N$ (I AM NOT SURE I UNDERSTAND THIS SENTENCE). 
This means that the phase sensitivity of the multi fringe approach operates at the SNL (like the two bin approach) as long as the number of fringes is adjusted according to the mean number of photons.   % see Supplementary material

For a fixed number of photons N, the number of fringes M depends on the choice of the parameter $b$ as well as on the required visibility. For a given set of N, M and $a$ there is an optimal value of $b$ that maximizes the visibility. Fig. ~3(c) reports two examples of the number of fringes as a function of the number of photons for two different visibilities. In this example, for each N, $b$ as been chosen such as to maximize M and to keep the visibility above 95\% or above 90\%. It is clear that at higher visibility the number of fringes for a given N decrease. The points in the figure are extracted by numerical calculationa and the curves are fits that scale as $M \propto \sqrt{N}$

%However, using multiple
%binning as opposed to binary, we achieve multiple fringes within
%one period. Indeed, when we divide the interval of $2\pi$ into $M$
%intervals $m=1,\ldots,M$ and define the detection operators
%$\hat{\Pi}_{0,m}$, $\hat{\Pi}_1$ in the same way as in (\ref{POVM}),
%we obtain a response structure with $M$ fringes per period,
%see Fig.~\ref{figure3} for $M=9\frac12$.
%Due to smaller distance between individual response peaks the
%overlap is higher and the visibility lower compared to the single
%fringe configuration. In fact, there is a trade-off between
%the number of fringes within a period and the visibility of
%the resulting pattern. Keeping the number of fringes linearly
%proportional to $\sqrt{N}$ the visibility remains constant
%independent of $N$. The sensitivity is multi fringe scheme
%reveals several minima, all of them follow the SNL scaling
%for increasing number of photons. % see Supplementary material

%\bigskip

In the above investigations we have considered only the measurement of the phase quadrature among all other quadratures. For the binary binning  approach, this is indeed the optimal quadrature measurement (for the input state considered in Fig.~\ref{figure1} (a)) although super resolution with reduced quality can be also obtained for any other quadrature measurement. This is clearly seen from the results of the multiple-bining approach in Fig. \ref{figure3} which shows that the fringes are narrowest at $\phi=0$ and broadest at $\phi=\pi/2$ effectively corresponding to a phase and an amplitude quadrature measurement, respectively. 
The fact that super resolution can be obtained for any quadrature measurement suggests that we may relax the stringent phase reference in our homodyne detector thus measuring a random quadrature and thereby attaining super resolution with a simplified strategy. 

%It is however also possible to achieve super resolution (although with reduced quality) and sensitivity at the SNL by measuring any other quadrature. This suggests that we may relax the stringent phase reference in our homodyne detector thus measuring a random quadrature and thereby attaining super resolution with a simplified strategy. 

%It can be clearly seen that local visibility of the pattern
%varies over the period. The scheme performs best for phases
%close to zero and worst for phase $\pm\pi/2$ which is equivalent
%to zero phase measured but HD set for $x$ projection instead
%of $p$. Even in this configuration the FWHM as well as the
%sensitivity of our method beat the parameters of a standard
%intensity detection system. Based on this observation our
%scheme can be extended to phase-averaged local oscillator
%yielding the visibility constant over the response pattern
%while keeping the improved resolution with $1/\sqrt{N}$
%scaling and sensitivity approaching the SNL arbitrarily.
%Releasing the precise phase reference condition enables
%us to compare our measurement device to theoretical proposals
%based on parity detection \cite{parity!!!}.
%Surprisingly, such highly non-classical detection scheme
%whose realization is out of reach of today state of the art
%photon counting technology gives the same resolution and
%sensitivity (up to a scaling constant) as what we presented
%experimentally here.

In contrast to previous super resolution schemes based
on NOON states or photon counters, the measurement presented
here is intrinsically deterministic. It means that we keep
every single measurement outcome and do not perform
a post selection of the outcomes to extract the desired super
resolving feature as done in previous implementations
\cite{Steinberg2004,Zeilinger2004,Takeuchi2007,White2007,Silberberg2010,Zappa2010}.
Due to this common post selection procedure (which significantly reduces the number
of available resource states), all these experiments exhibit a phase sensitivity that is lower than
the one obtained here if the actual number of photons passing
through the phase sample is taken into acccount.
%Likewise, the resolution of previous implementations
%is practically limited by the number of entangled photons
%in the NOON state or by the number of photons that can be resolved
%in a photon counter, both of which are presently limited to approximately five.
%Using the scheme outlined here the resolution limit can
%in practice be arbitrarily small.
%For example, using $1$~pW of visible
%laser radiation, 
%we find a FWHM phase resolution of the order of $0.1$~nm.
%A similar resolution can be obtained by using an experimentally unfeasible
%$500$ photon NOON state. Moreover, using such a state in the presence
%of just $0.2\%$ loss, it is impossible to attain sensitivity operation
%at the SNL due to the extreme fragility of high NOON states \cite{Rubin2007}.
%In contrast, the performance of our method is basically unaltered in the presence of loss since the loss can be ideally compensated by increasing the mean number of photon of the input state. 

%By using a squeezed state in replacement of the coherent state
%in the interferometer, we have found theoretically that the
%interferometric scheme with a dichotomic homodyne detector will
%in addition to the super resolving capabilities exhibit a phase
%sensitivity that beats the SNL. Such a scheme
%will be investigated in the future. 

Using a very simple setup based on coherent states and a high-efficiency homodyne detector, we have demonstrated a narrowing of the interference fringes of an interferometer beyond what is possible with conventional interferometers. In contrast to previous implementations of super-resolution with coherent states, the proposed scheme is deterministic and it attains a phase-sensitivty at the shot noise limit. Both the phase resolution and the phase sensitivity was found to scale inversely with the coherent state amplitude.  

{\it  Acknowledgement:}
The work was financed by the Danish Agency for Science, Technology and Innovation (Sapere Aude).
MJ also acknowledges the support by Projects No. MSM6198959213
and No. LC06007 of the Czech Ministry of Education.
We thank M. Paris, J. Fiurasek, R. Filip and A. Tipsmark for valuable discussions.

%Another area for future
%research is to study the super resolving scheme in a multimode
%configuration combined with multi-pixel detectors to enhance
%the resolution of phase images.

% In setting up this template for *Science* papers, we've used both
% the \section* command and the \paragraph* command for topical
% divisions.  Which you use will of course depend on the type of paper
% you're writing.  Review Articles tend to have displayed headings, for
% which \section* is more appropriate; Research Articles, when they have
% formal topical divisions at all, tend to signal them with bold text
% that runs into the paragraph, for which \paragraph* is the right
% choice.  Either way, use the asterisk (*) modifier, as shown, to

%\begin{acknowledgment}
%\end{acknowledgment}

%%%%%%%%%%%%%%
%% FIGURES %%%
%%%%%%%%%%%%%%

% \begin{figure}
% 	\centering	
% 	\includegraphics[width=14cm]{./imperfect_diff_a_n100_paper.pdf}
% 	\caption{The theoretical response function $f_a(\phi)$ for mean number
%of photons $N=100$ and different values of $a$. $a=0.1$, dashed black line,
%$a=0.5$, red, $a=1$, blue. The dotted line represents $\mathrm{cos}^2(\phi/2)$,
%the typical response function for a Rayleigh limited protocol.}  
% \end{figure}


\begin{thebibliography}{99}

% classical introduction to optics + Rayleigh

% beating Rayleigh limit with tomography and maxlik reconstruction
%\bibitem{Jezek2004}
%Je\v{z}ek, M. and Hradil, Z.
%Reconstruction of spatial, phase and coherence properties of light.
%J. Opt. Soc. Am. A 21, 081407 (2004).

% beating Rayleigh limit by non-linear interaction with matter
% \bibitem{STED}
% Dyba, M. and Hell, S. W. Focal spots of size λ/23 open
% up far-field fluorescence microscopy at 33 nm axial resolution.
% Phys. Rev. Lett. 88, 163901 (2002).
% Rittweger, E. et al. STED microscopy reveals crystal colour centres
% with nanometric resolution. Nature Photonics 3, 144-147 (2009).

% quantum litography using NOON

\bibitem{BornWolf}
Born, M., Wolf, E. Principle of Optics. Cambridge University Press, 1999.

\bibitem{Boto2000}
Boto, A.N. et al.
Quantum interferometric optical litography: exploiting entanglement
to beat the diffraction limit.
Phys. Rev. Lett. 85, 2733-2736 (2000).

\bibitem{Wineland2005}
Leibfried, D. et al.
Creation of a six-atom Schr\"odinger cat state.
Nature 438, 639-642 (2005).
\bibitem{Morton2009}
Jones, J.A. et al.
Magnetic field sensing beyond the standard quantum limit using 10-spin NOON states.
Science 324, 1166-1168 (2009).
\bibitem{Chen2010} 
Chen, Y.-A. et al. 
Heralded Generation of an Atomic NOON State
Phys. Rev. Lett, 104, 043601 (2010).
\bibitem{Rarity1990} Rarity, J. G. et al. Phys. Rev. Lett. 65, 1348 (1990).
\bibitem{Steinberg2004}
Mitchell, M.W., Lundeen, J.S. and Steinberg, A.M.
Super-resolving phase measurements with a multiphoton entangled state.
Nature 429, 161-164 (2004).
\bibitem{Zeilinger2004}
Walther, P. et al
De Broglie wavelength of a non-local four-photon state.
Nature 429, 158-161 (2004).
\bibitem{Takeuchi2007}
Nagata, T. et al.
Beating the standard quantum limit with four-entangled photons.
Science 316, 726-729 (2007).
\bibitem{giovanetti2011} 
Giovanetti, V., Lloyd, S., Maccone, L. Nat. Photon. 5, 222 (2011).
\bibitem{Fiurasek2002}
Fiur\'a\v{s}ek, J. Conditional generation of $N$-photon entangled states of light.
Phys. Rev. A 65, 053818 (2002).
% \bibitem{Toll1956}
% Toll, J.S. Causality and the dispersion relation: logical foundations.
% Phys. Rev. 104, 1760-1770 (1956).
\bibitem{Rubin2007}
Rubin, M.A. and Kaushik, S.
Loss-induced limits to phase measurement precision with maximally entangled states.
Phys. Rev. A 75, 053805 (2007).
\bibitem{Walmsley2009}
Dorner, U. et al. Optimal Quantum Phase Estimation.
Phys. Rev. Lett. 102, 040403 (2009).
%Banaszek, K. et al. Quantum states made to measure.
%Nature Photonics 3, 673-676 (2009).
% time reversal approach, projection to NOON states and parity det
\bibitem{Pregnell2004} Pregnell, K.L. and Pegg, D.T.
J. Mod. Opt. 51, 1613 (2004).
\bibitem{Boyd2004}
Bentley S.J. and Boyd, R.W.
Nonlinear optical lithography with ultra-high sub-Rayleigh resolution.
Opt. Express 12, 5735, (2004).
\bibitem{Bouwmeester2006}
Khoury G. et al.
Nonlinear Interferometry via Fock State Projection.
Phys. Rev. Lett. 96, 203601 (2006).
\bibitem{White2007}
Resch, K.J. et al.
Time-reversal and super-resolving phase measurement.
Phys. Rev. Lett. 98, 223601 (2007).
\bibitem{Silberberg2010}
Afek, I., Ambar, O. and Silberberg, Y.
High-NOON states by mixing quantum and classical light.
Science 328, 879-881 (2010).
\bibitem{Dowling2010}
Gao, Y.  et al.
Super-resolution at the shot noise limit with coherent states and photon-number-resolving detectors.
J. Opt. Soc. Am. B 27, A170 (2010).
\bibitem{Zappa2010}
Guerrieri F. et al.
Sub-Rayleigh Imaging via N photon detection.
Phys. Rev. Lett.  105, 163602 (2010).
\bibitem{Kothe2010}
Kothe C., Bjork G., Bourennane M. 
Arbitrary high super-resolving phase measurements at telecommunication wavelengths
Physical Review A 81, 063836 (2010)
\bibitem{Boyd} 
Shin, H., Chan, K.-W.-C., Chang, H.J. and Boyd, R.W.
Phys. Rev. Lett. 107, 083603 (2011).
\bibitem{Plick2010}
Plick William, N. et al.
New J. Phys. 12, 113025 (2010).



%\bibitem{Tsang} 
%Tsang, M. 
%Phys. Rev. Lett. 102, 253601 (2009). 
%\bibitem{Walmsley} 
%Thomas-Peter, N., Smith, B., Datta, A., Zhang, L., Dorner, U., and Walmsley, A.
%Phys. Rev. Lett. 107, 113603 (2011). 

% \bibitem{Okamoto2008}
% Okamoto, R. et al.
% Beating the standard quantum limit: phase super-sensitivity of N-photon interferometers.
% New J. Phys. 10, 073033 (2008).

% NOON preparation efficiency and susceptibility to losses



%\bibitem{Giovannetti2006}
%Giovannetti, V. et al. Quantum Metrology.
%Phys. Rev. Lett. 96, 010401 (2006).


\end{thebibliography}
\end{document}